\documentclass[5p,times]{elsarticle}
\usepackage{amssymb}
\usepackage{amsmath}
\usepackage[hyphens]{url}
\usepackage{enumitem}
\usepackage[frozencache,cachedir=.]{minted}
\usepackage{algorithm}
\usepackage{algpseudocode}
\usepackage{array, booktabs, makecell}
\usepackage[lofdepth,lotdepth]{subfig}
\usepackage[table,xcdraw]{xcolor}
\begin{document}

\begin{frontmatter}
\title{Implementation of an efficient, portable and platform-agnostic cryptocurrency mining algorithm for Internet of Things devices}

\author[inst1]{Kinshuk Dua}
\ead{kinshukdua@gmail.com}
\affiliation[inst1]{organization={School of Computer Science and Engineering, Vellore Institute of Technology},
            city={Chennai},
            state={Tamil Nadu},
            country={India}}

\begin{abstract}

Recently, there has been a remarkable amount of research being done in both, the fields of Blockchain and Internet of Things (IoT). Blockchain technology synergises well with IoT, solving key problems such as privacy, concerns with interoperability and security. However, the consensus mechanisms that allows trustless parties to maintain an agreement, the same algorithms that underpins cryptocurrency mining, are usually extremely computationally expensive, making implementation on low-power IoT devices difficult. More importantly, mining requires downloading and synchronizing hundred of gigabytes worth of blocks which is far beyond the capabilities of most IoT devices. In this paper, we present an efficient, portable and platform-agnostic cryptocurrency mining algorithm using the Stratum protocol to avoid downloading the entire blockchain. We implement the algorithm in four different platforms- PC, ESP32, an emulator and an old PlayStation Portable (PSP) to demonstrate that it is indeed possible for any device to mine cryptocurrencies with no assumptions except the ability to connect to the internet. To make sure of ease of portability on any platform and for reproducibility of the reported results we make the implementation publicly available with detailed instructions at: https://anonymous.4open.science/r/cryptominer.

\end{abstract}

\end{frontmatter}

\section{Introduction}
\label{sec:introduction}
The concept of a decentralized, trustless digital currency was first described in 2008 by Satoshi Nakamoto, while the source code of the first implementation was released under an open source license in 2009 \cite{nakamoto2008bitcoin,chohan2017history}. Even though over 6000 cryptocurrencies \cite{coinmarketcap} have been developed, Bitcoin firmly holds the ground as the most used cryptocurrency. 
Bitcoin uses a proof of work system to implement a distributed timestamp server on a peer-to-peer basis. The distributed timestamp solves the core issue of double-spending on a distributed ledger with no central authority. 
A fundamental concept of Bitcoin and other similar cryptocurrencies that use proof-of-work consensus mechanism is that "work", using computational power, required for proof of work, must be done to ensure the reliability of the blockchain which in turn creates the cryptocurrency as a result. This is known as mining and the devices that mine the cryptocurrencies are known as miners. The first block of a network is known as the genesis block. In case of bitcoin it was mined by Nakamoto who received 50 bitcoins as a reward, this represented the first ever Bitcoin or cryptocurrency transaction \cite{o2014bitcoin}. 
The SHA-256 cryptographic hash function is used in the proof of work system to generate the hash digest used for validating the transactions and the blockchain \cite{nakamoto2008bitcoin}. The SHA-256 function is virtually found in all modern devices that connect to the internet because a majority of the websites use HTTPS (Hypertext Transfer Protocol Secure) which uses SSL/TLS which in turn requires the SHA-256 cryptographic hash function. 
A lot of old and low cost embedded IoT systems lack the necessary pre-requisites (such as the C standard library) required to work with existing SHA-256 libraries or do not have the ability to access the built in SHA-256 function used for the SSL certificate. Furthermore, the libraries are developed to be cross-platform which means that optimizations for the specific microproccessor used in the embedded system cannot be implemented. Moreover, most IoT devices lack high capacity storage needed to store the entire blockchain.
This paper proposes a SHA-256 based cryptocurrency mining algorithm using the Stratum protocol used for combining the processing power of multiple miners. Since Bitcoin is the most popular cryptocurrency with the most number of public pools available it was chosen as the cryptocurrency to be mined and the rest of the paper focuses on it, however the algorithm is portable and can be used to mine any cryptocurrency with relatively little change. For the implementation, we chose 4 different platforms- an x64 PC, a PSP emulator, an old embedded device (Sony's PlayStation Portable) as it is low powered, obsolete and has minimal networking capabilities and finally an ESP32 a new low-cost IoT device to show that any computing device capable of connecting to the internet (including IoT devices) can be converted to a cryptocurrency miner. 

\section{Motivation}

Cryptocurrencies have enjoyed unprecedented growth in the past few years, more than one in ten Americans invested into cryptocurrencies in the last year \cite{norc2021crypto} and around 80\% of people in Asia were aware of cryptocurrencies in 2019 \cite{oecd2019survey}. Recently, China has launched a national cryptocurrency and the country of El Salvador has even adopted Bitcoin as a legal tender. As we move into a more digitized world, cryptocurrencies will become even more widespread.

For cryptocurrencies that depend on the proof-of-work consensus mechanism like Bitcoin, mining is of utmost importance as it guarantees security, prevents tempering, confirms transactions and generates new currencies (Section \ref{sec:bitcoinalgo}). However, mining is a computationally expensive task and usually requires dedicated hardware. Moreover, the entire blockchain needs to be replicated on all the nodes which requires hundreds of GBs of storage and is therefore, not currently feasible or possible in low powered IoT devices. We summarize our contributions below-
\begin{enumerate}
    \item To show that it is possible to run a mining algorithm on any device connected to the internet. By showcasing an efficient implementation of the mining algorithm on a device not meant for cryptocurrency mining, We show that it is possible to harness the computing power that might otherwise go to waste. We can cluster multiple such miners running on heterogeneous clients and then substitute them in place of a high powered dedicated device. The average person generates about 7.3 Kgs of e-waste in an year, a lot of the times discarding perfectly functional devices \cite{forti2020global}. By reusing obsolete or unused devices not only can we reduce the overall e-waste generated but also give people an incentive to do so. 
    
    \item To showcase an algorithm for implementing consensus mechanisms for blockchains on resource-constrained devices. Blockchain solves a lot of problems faced by IoT devices like privacy concerns and inadequate interoperability by providing privacy, transparency, reliability and security \cite{dai2019blockchain}, however as noted above, it is difficult or outright impossible to store the whole blockchain on the device thus making many solutions purely theoretical. We show that it is indeed possible to implement blockchains on IoT devices even if they don't have enough storage or built-in protocols for mining.
    
    \item To improve security of small and/or private networks. Private or small blockchain networks can benefit hugely from more devices and thus more computing power as addition of more devices and thus more computing power increases the security of the whole network. The algorithm can even also be modified to run other internet based distributed computations that require consensus mechanisms.
    
    \item To introduce the possibility that cryptocurrency mining malware might target IoT devices seeing that there is now a financial motive to do so. Especially since, IoT devices are known to be insecure and vulnerable to attacks\cite{khan2018iot}. 
    
    \item When Proof of Stake (PoS) becomes commonplace it might be worthwhile to use algorithm similar to ours on low-power devices to restrict electricity cost.
\end{enumerate}

\section{Brief Review of Bitcoin mining}

Because of the open source nature of Bitcoin, thousands of new alternative cryptocurrencies have since popped up over the years, each with its own separate client, consensus algorithm, platform and mechanisms. In this paper however, the focus will only be on cryptocurrencies that use the SHA-256 proof of work consensus mechanism. Even though only the bitcoin mining algorithm is described here in detail, the algorithm can be trivially modified to support most other cryptocurrencies which use the SHA-256 proof of work algorithm.

\subsection{The Bitcoin mining algorithm}
\label{sec:bitcoinalgo}
Each bitcoin is defined as a chain of digital signatures in the form of transactions. The holder of the coin can transfer the currency by digitally signing a hash of the last transaction along with public key of the future owner of the coin and appending the same to the end of the current chain of signatures. To verify the legitimacy of the coin, anyone can verify the chain of signatures since all transactions are public in bitcoin. 

To ensure that the previous owner of the coin didn't already transfer the coin in an earlier transaction (and thus to prevent double spending) the hash of the block along with the timestamp is published to all the nodes of the bitcoin network. Each timestamp includes all the previous timestamps to reinforce the ones before it.

To implement the distributed timestamp server on the peer-to-peer network, bitcoin uses the hashcash \cite{back2002hashcash} proof of work algorithm with SHA-256 being used as the cryptographic hash function. The algorithm is a complex cryptographic puzzle with the goal of getting the hash value less than the given target value. First the timestamp and the hash of the previous block is appended to the end of the transaction root and hashed. Then a special number called nonce (short for "number only used once") is appended to the newly generated hash and this whole number is again hashed. This final hash digest is compared with the special number called the target value. If the hash is less than or equal to the target then the block is said to be mined and the miner is rewarded with a fixed number of bitcoins. The target is a 256-bit number that all miners on the networks share, the target has a specific number of zeroes in front of it. Since the SHA-256  is a cryptographic hash function, the only way to make sure that the hash is less than or equal to the target is to brute force the nonce, which means the average work required is exponential with respect to the leading zeroes in the target. Although other methods instead of brute-forcing exist such as the SAT solving \cite{courtois2014optimizing}, the average work required will always remain exponential.
To compensate for increasing hardware speed and varying interest in running nodes over time, the proof-of-work target is adjusted, it is determined by a moving average targeting an average number of blocks per hour. If the blocks are being mined too fast, the target value is decreased \cite{nakamoto2008bitcoin}.

\subsection{The original Bitcoin client}
The first implementation of the bitcoin architecture described by Satoshi Nakamoto was in the form of the open-source bitcoin-client released in 2009, which used the OpenSSL library to Implement the SHA-256 algorithm.  The OpenSSL library is available as an open-source software with official build instructions for Windows, Android, DOS, OpenVMS and Unix-like platforms. However the library has to be manually cross compiled to the desired platform for it to work on a new hardware architecture such as the  MIPS microprocessor used in this paper and even still there is no way to ensure compatibility without painstakingly going through and rewriting thousands of lines of codes. Starting with version 0.3.22 the ability to mine coins on the original bitcoin client was removed \cite{bitcoinv0322}. This was done because at the time specialized mining clients were getting much more efficient, especially on dedicated hardware such as GPUs, compared to the CPU mining capabilities that the original client had \cite{skudnov2012bitcoin}.

\subsection{GPU mining}
Since mining consists of applying the same hash function while iterating over the the nonce, the task is embarrassingly parallel \cite{vilim2016approximate}. We can use methods like loop-unrolling and take advantage of SIMD execution and run the computation in parallel on thousands of cores found in modern GPUs. 

In less than an year after the release of the first bitcoin client, the mining capabilities found on the bitcoin client was rendered obsolete because at the time the GPU miners were over 100 times more efficient than the CPU miner found on the original client \cite{skudnov2012bitcoin}. Presently even GPU mining is obsolete since dedicated hardware like Field-programmable gate array (FPGA) and later Application-specific integrated circuit (ASIC) took over as the most efficient (and only profitable) way to mine cryptocurrencies.

\subsection{Pool Mining}

\begin{figure}[h]
\caption{The plot of the actual bitcoin network difficulty over time}
\centering
\includegraphics[width=8cm]{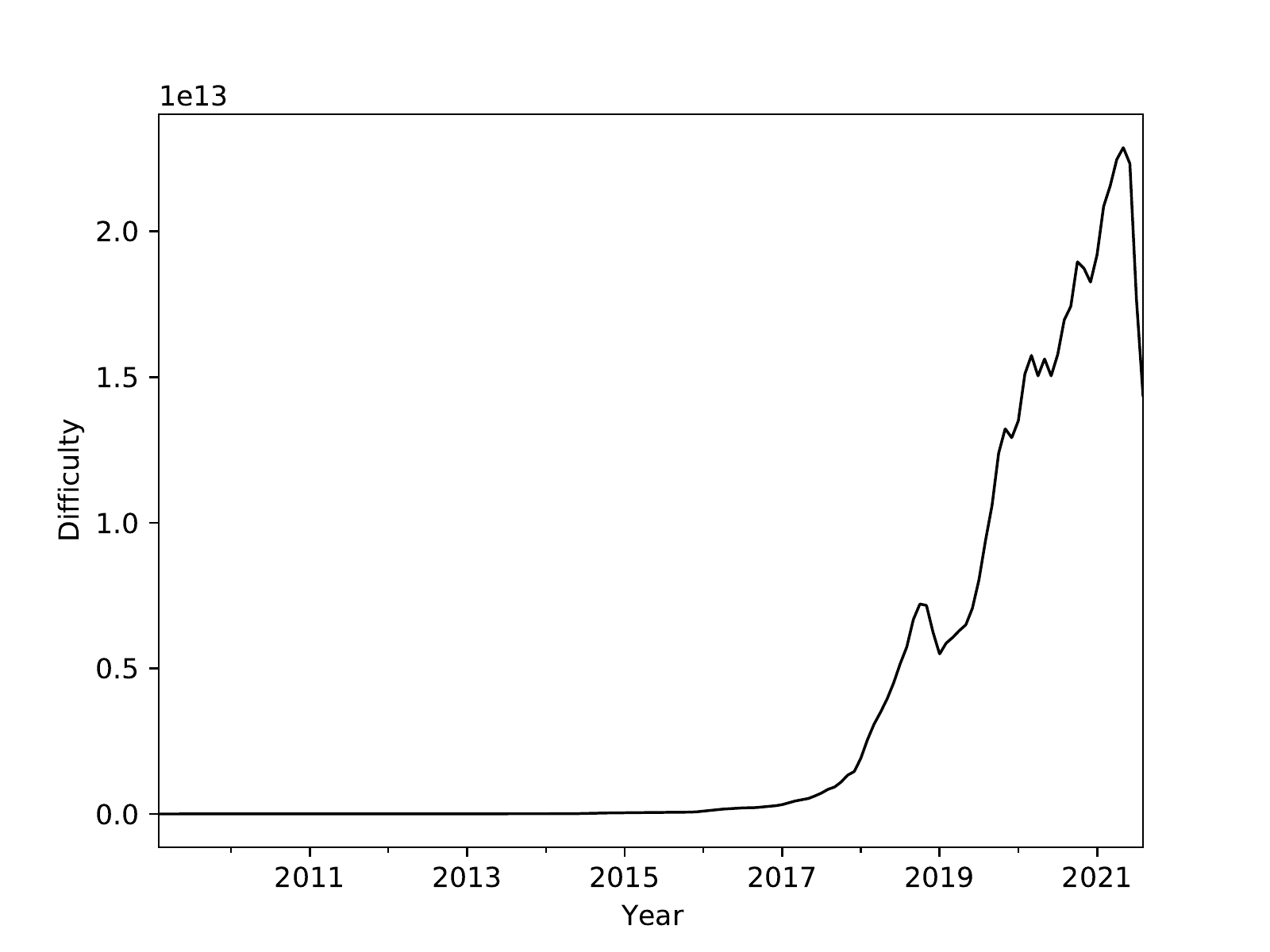}
\label{fig:diff_rate}
\end{figure}

\begin{figure}[h]
\centering
\includegraphics[width=8cm]{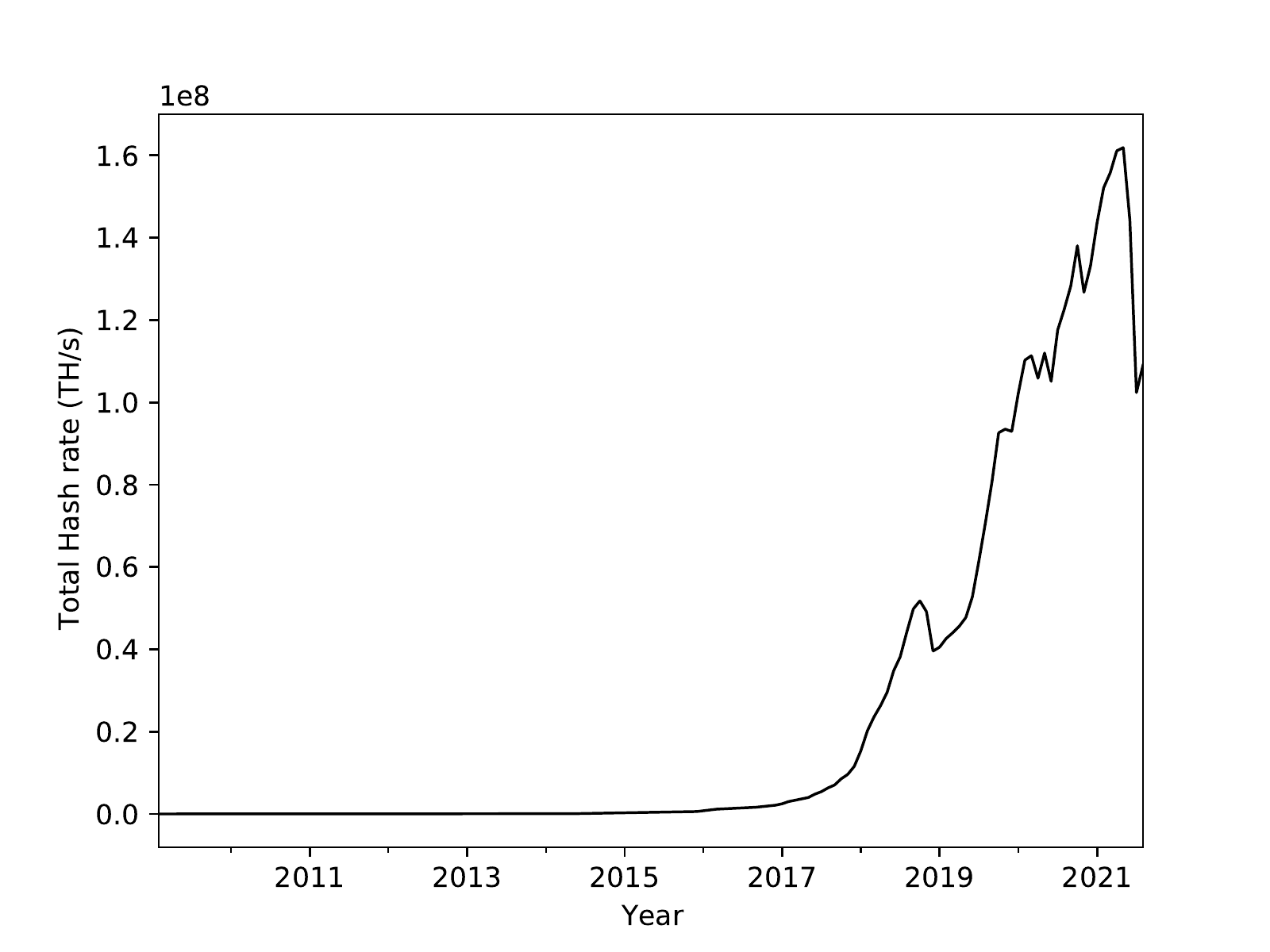}
\caption{Plot of the Total Network Hashrate (in TH/s) over time.}
\label{fig:hash_rate}
\end{figure}

\begin{figure}[h]
\includegraphics[width=8cm]{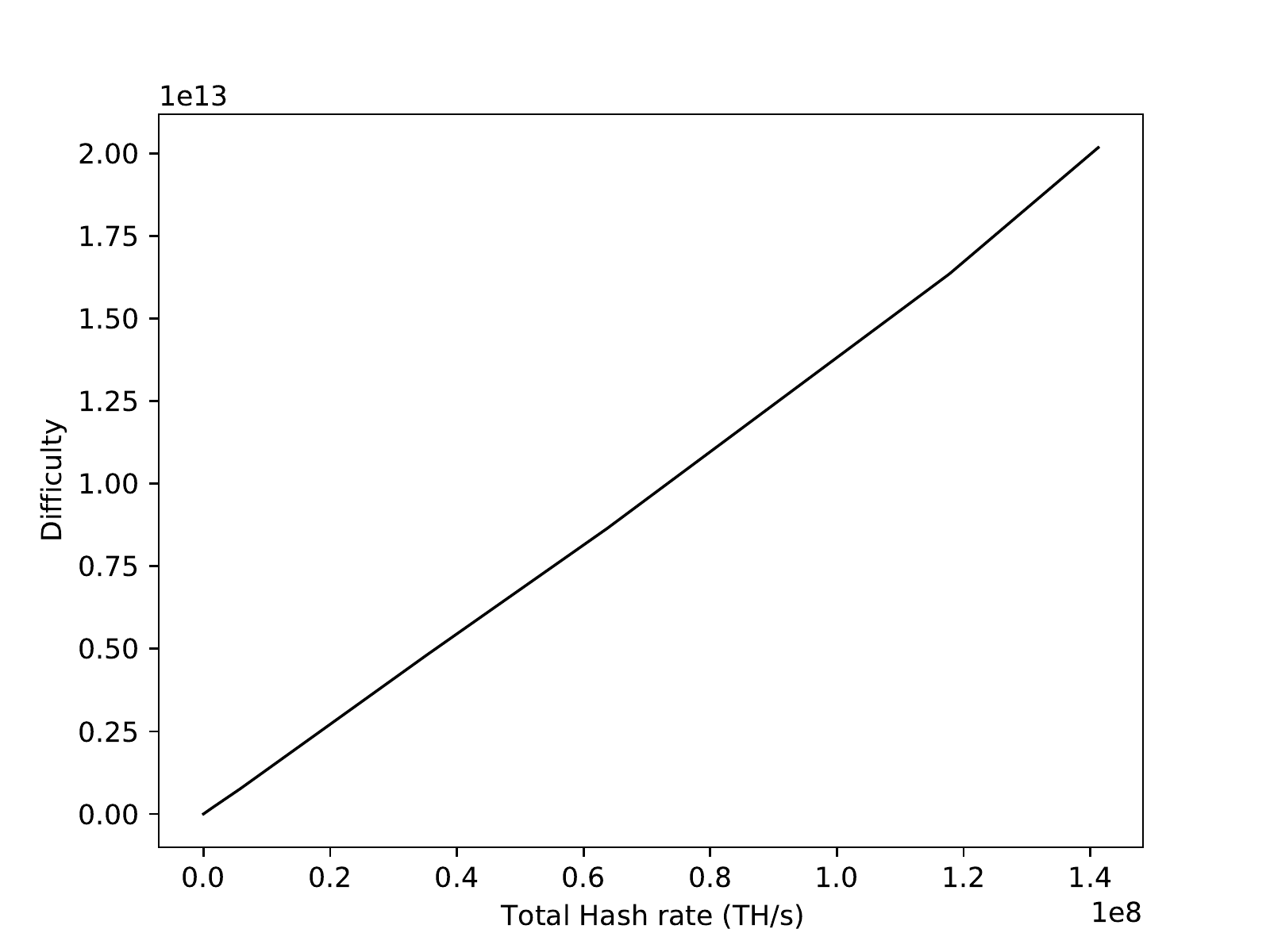}
\caption{Plot of Difficulty and Hashrate (in TH/s) based on data from 2009-2021}
\label{fig:diff_vs_hash}
\end{figure}

As mining became increasingly mainstream, the difficulty of the bitcoin network increased exponentially (Figure \ref{fig:diff_rate}). This was caused by massive increase in the total hash rate of the network (Figure \ref{fig:hash_rate}). Both the graphs look almost identical, if we compare the total hash rate and the difficulty we can see that the difficulty is directly proportional to the total hash rate of the network (Figure \ref{fig:diff_vs_hash}). 

The difficulty increases every 2016 blocks. The target is related to the difficulty with the formula 

\begin{displaymath}
\text{difficulty}= \frac{\text{difficulty\_1\_target}}{\text{target}}
\end{displaymath}

Where `$\text{difficulty\_1\_target}$` is the target value of the genesis block.

Even with expensive, extremely powerful and efficient devices it takes a very long time to generate a block. For example, with the current difficulty of $13912524048946$ it will take the fastest commercially available ASIC miner with a hashrate of 100 TH/s around 19 years (Eq. \ref{form:mineone}) to mine one block.
\begin{equation}
\label{form:mineone}
\frac{13912524048946 \times 2^{32}}{100 \cdot 10^{12} \times 24 \times 3600 \times 365} \approx  18.95 \text{ years}
\end{equation}
To fix the volatility of the block rewards, user \textit{slush} on a bitcoin forum devised a mechanism for users to cluster their resources because there was an overlap in hash calculations done by different users which led to wastage of computing power- "...everybody counts sha256 hashes from the same range. Two separate CPUs with 1000khash/s isn't the same as one 2000khash/s machine!". This led to the creation of the first Bitcoin mining pool \cite{poolmine}. 

Even though the time to mine a block remains constant ($\approx 10$ mins), the whole network is trying to mine that same block, so the probability of a single miner, especially one with a low hashrate, being successful is miniscule. 

Soon thereafter pools became the primary way to mine bitcoins, since mining pools guaranteed smaller but more frequent rewards and thus were less volatile and more profitable. Currently, mining pools account for an astonishing 93.2\% of all mining hashrate in the network \cite{poolshare}.

\subsubsection{Stratum Protocol}

\begin{figure}[h]
\centering
\includegraphics[width=0.45\textwidth]{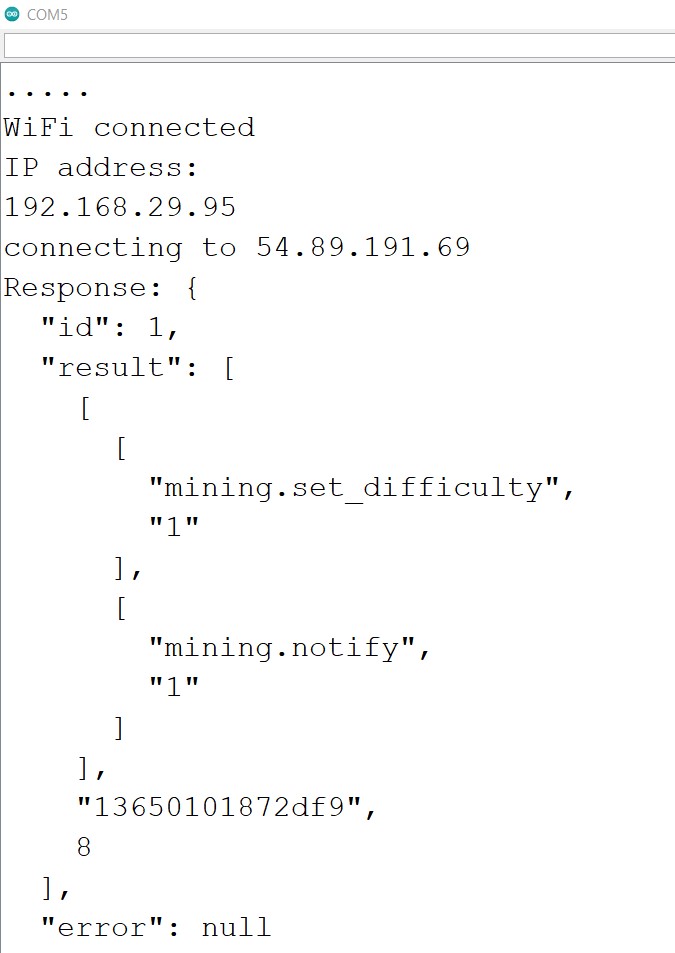}
\caption{Stratum protocol on ESP32}
\label{fig:response}
\end{figure}

Most early pools worked on the `getwork` protocol implemented in the official bitcoin client \cite{ skudnov2012bitcoin, poolmine}. In 2012 in order to solve the drawbacks of the then widely used `getwork` protocol, Stratum was introduced. Stratum is a line-based protocol implemented with TCP sockets, it uses JSON-RPC (JavaScript Object Notation - Remote Procedure Call) to encode the messages and provide a efficient way to communicate between the server and mining devices. It offers better performance because it is based on TCP in contrast to getwork which was based on HTTP and had the associates overheads and uses long polling. Long polling prevents idling of miners on network latency and scales much better especially with fast miners \cite{stratumdocs}.
Even though Stratum is widely used it lacks a formal Bitcoin Improvement Proposal (BIP) and hence has no official specification  \cite{slushstratum}.

The Stratum Protocol has the following methods (each method must be followed by a line break) :
\begin{enumerate}[label=(\alph*)] 
    \item Client methods
\begin{enumerate}[label=\arabic*.]

    \item mining.subscribe - subscribes the client for mining job notifications
        \begin{minted}[breaklines,frame=single,fontsize=\footnotesize]{json}
{"id": 1, "method": "mining.subscribe", "params": ["User agent/Empty string", "session id", "host pool address", "port"]}
        \end{minted}
        Response:
        \begin{minted}[breaklines,frame=single,fontsize=\footnotesize]{json}
{"id": 1, "result": [ [ ["mining.set_difficulty", "subscription id 1"], ["mining.notify", "subscription id 2"]], "extranonce2", "size of extranonce2"], "error": null}
        \end{minted}

    \item mining.authorize - authorizes the client with the given credentials
    \begin{minted}[breaklines,frame=single,fontsize=\footnotesize]{json}
{"id": 2, "method": "mining.authorize", "params": ["username", "password"]}
    \end{minted}
    Response:
        \begin{minted}[breaklines,frame=single,fontsize=\footnotesize]{json}
{"error": null, "id": 2, "result": true}
        \end{minted}

    \item mining.submit - submits the solution for the job
    \begin{minted}[breaklines,frame=single,fontsize=\footnotesize]{json}
{"params": ["username", "jobid", "extranonce2", "ntime", "nonce"], "id": 3, "method": "mining.submit"}
    \end{minted}

    Response:
        \begin{minted}[breaklines,frame=single,fontsize=\footnotesize]{json}
{"error": null, "id": 3, "result": true}
        \end{minted}
        OR
         \begin{minted}[breaklines,frame=single,fontsize=\footnotesize]{json}
{"id": 3, "result": null, "error": [21, "Job not found", null]}
        \end{minted}      
\end{enumerate}
    \item Server methods
\begin{enumerate}[label=\arabic*.]

    \item mining.notify - notifies the workers about the job
        \begin{minted}[breaklines,frame=single,fontsize=\footnotesize]{json}
{"id": null, "method": "mining.notify", "params": ["job_id", "previous hash",
"start of coinbase",
"end of coinbase", ["merkle branch"],
"version", "encoded network difficulty", "ntime", false]}
        \end{minted}
        
        \item mining.set\_target - sets the difficulty for all the worksers from the next job onwards
        \begin{minted}[breaklines,frame=single,fontsize=\footnotesize]{json}
{ "id": null, "method": "mining.set_difficulty", "params": [2]}
        \end{minted}
\end{enumerate}
\end{enumerate}
\section{Implementation}
\subsection{Hardware Specification}
The hashing algorithm is implemented on a variant of the MIPS R4000 microprocessor and the communication via the Stratum protocol is done using the Wi-Fi IEEE 802.11b compatible ARM9 processor.

\subsubsection*{MIPS R4000}
The R4000 is a 64-bit, highly integrated RISC microprocessor based on the Microprocessor without Interlocked Pipelined Stages (MIPS) Instruction Set Architecture with support for superpipelining, floating point instructions \cite{mirapuri1992mips}.
The variant used in this paper had a clock speed of 20-333 Mhz.

\subsubsection*{PlayStation Portable}
The Sony PlayStation Portable (PSP) was a popular handheld game console. Originally released in 2004, the PSP was equipped with the aforementioned R4000-based processor, 32 Megabytes of RAM (4 Megabytes of embedded DRAM), 4.2 inch TFT LCD with 480×272 pixel resolution and an ARM9 Processor with 802.11b Wi-Fi connectivity. In 2007, PSP-2000, a revised version of the PSP was released which was thinner and lighter but more importantly had double the system RAM (64 Megabytes) \cite{hachman_2004,conrad2009forensic}. The PSP supports multiple threads, but because of the presence of only one core, it offers no practical improvement \cite{pspthread}. 

\subsubsection*{ESP32}
ESP32 is a low-cost, low-power, general purpose,  microprocessor with integrated Bluetooth and Wi-Fi and no inbuilt storage. It comes with a Xtensa dual-core 32-bit LX6 microprocessor, operating at 160 or 240 MHz and 320 KiB RAM, making it perfect for implementing the algorithm.

\subsection{SHA-256}
Two implementations of the SHA-256 algorithm were tested
\begin{enumerate}
    \item A naive implementation adapted from \cite{sha2naive}.
    \item A heavily optimised implementation adapted from CGminer \cite{sha2cgminer}.
\end{enumerate}

\subsection{Implementation Details}
The code is largely self contained requiring no additional libraries aside from the source code to make it truly portable and platform independent. Only some low level functionalities have to provided by the underlying platform. 
\begin{enumerate}
    \item Ability to exchange TCP messages
    \item Run-time memory allocation
    \item Basic Memory management 
    (for concatenating bytes and comparing string)
\end{enumerate}
Since almost all IoT and embedded devices have these basic abilities, we can run the proposed cryptocurrency mining algorithm on them.
\subsection{Algorithm}

The algorithm consists of the following steps:
\begin{enumerate}
    \item Connect to the mining pool and get mining jobs.
            \begin{enumerate}[label=\roman*.] 
            \item Connect to the Stratum pool server using a TCP Socket.
            \item Call the mining.subscribe method to initialize a connection with the pool server. \\
            The server will return:
            \begin{itemize}
                \item Subscription ID for mining.set\_difficulty.
                \item Subscription ID for mining.notify.
                \item Hexcoded extranonce1.
                \item Size of extranonce2.
            \end{itemize}
            \item Authorize the worker with mining.authorize method. \\
            The result of the authorization is returned.
            \item Almost immediately after subscription, the miner will start receiving mining job. \\
            A mining job consists of: 
            \begin{itemize}
                \item Job ID - Used for submitting the job.
                \item Previous hash.
                \item Initial part of the coinbase.
                \item Final part of the coinbase.
                \item Merkle branches - a list of pre-prepared hashes of the transactions for generating merkle root.
                \item Encoded current network difficulty.
                \item ntime - Current time in Unix time format.
                \item Block version.
                \item Whether to discard all previous jobs. 
            \end{itemize}
        \end{enumerate}
    \item \label{block:start} Generate the blockheader for each job.
        \begin{enumerate}[label=\roman*.]
            \item  Generate a unique extranonce2 (of appropriate size) for the given job.
            \item Build the coinbase transaction by concatenating initial part of the coinbase, extranonce1, extranonce2, and final part of the coinbase.
            \item Apply double SHA-256 on the coinbase transaction.
            \item Set the coinbase transaction's hash as the merkle root.
            \item For each hash in the list of pre-prepared hashes, append the hash to the merkle root and take double SHA-256. Set the output of the double SHA-256 digest as the new merkle root and repeat.
            \item Concatenate block version, previous hash, newly generated merkle root in reverse order, "current" time, encoded network difficulty, ntime and store it as the blockheader.
        \end{enumerate}

    \item Mine a block by iterating the nonce.
        \begin{enumerate}[label=\roman*.]
            \item \label{mine:start} Initialize the 32 bit nonce with zero.
            \item Append the nonce to the block header.
            \item Take the double SHA-256 of the block header.
            \item \label{mine:check_target} If the hash is smaller than or equal to the target, submit the job along with the winning nonce and extranonce2 to the server using the mining.submit command. \\
            The status of the submitted job is returned (accepted or rejected).
            \item Otherwise increment the nonce and repeat steps \ref{mine:start} through \ref{mine:check_target}.
            \item If the nonce overflows, increment the extranonce2 and restart from step \ref{block:start} to generate a new block header.
        \end{enumerate}
\end{enumerate}
For the complete algorithm refer to Algorithm \ref{alg:main}.

\MakeRobust{\Call}
\algnewcommand\True{\textbf{true}\space}
\algnewcommand{\LineComment}[1]{\State \(\triangleright\) #1}
\algnewcommand\algorithmicforeach{\textbf{for each}}
\algdef{S}[FOR]{ForEach}[1]{\algorithmicforeach\ #1\ \algorithmicdo}
\newcommand*\BitAnd{\mathbin{\&}}
\newcommand*\BitOr{\mathbin{|}}
\newcommand*\ShiftLeft{\ll}
\newcommand*\ShiftRight{\gg}
\newcommand*\BitNeg{\ensuremath{\mathord{\sim}}}
\newcommand{\algorithmicbreak}{\textbf{break}}
\newcommand{\Break}{\State \algorithmicbreak}

\begin{algorithm}
\caption{The mining algorithm}\label{alg:main}
\begin{algorithmic}
\LineComment{ + is concatenation operator}
\LineComment{ ++ is the increment operator}
\LineComment{ $\BitAnd$ is the Bitwise AND operator}
\LineComment{ $\ShiftRight$ is the Bitwise Shift Right operator}
\LineComment{ $\times$ is the multiplication operator}
\LineComment{ $-$ is the subtraction operator}
\LineComment{ $0$x implies that the number is represented in the Hexadecimal notation}

\Function{getTarget}{nCompact}
    \State nSize $\gets$ nCompact $\ShiftRight$ 24
    \State nWord $\gets$ nCompact $\BitAnd$ 0x007fffff
    \If{nSize $\leq 3$}
        \State nWord $\gets$ nWord $\ShiftRight ( 8 \times ( 3 -$ nSize$))$
    \Else
        \State nWord $\gets$ nWord $\ShiftRight ( 8 \times ($ nSize $-3))$
    \EndIf
    \State \Return nWord
\EndFunction
\Function{doubleSHA2}{message}
    \State \Return $\Call{SHA256}{\Call{SHA256}{message}}$
\EndFunction
\State sock $\gets \Call{TCP\_socket.connect}{server\_address, port}$
\State JobQueue
\State $\Call{mining.subscribe}{Useragent, session\_id, server\_address, port}$
\State sid1, sid2, extranonce1, extranonce2\_size $\gets \Call{sock.listen}{\null}$ 
\State $\Call{mining.authorize}{username,password}$
\State jobid, prevhash, coin1, coin2, merkle\_branch, nbits, ntime, ver, discard $\gets \Call{sock.listen}{\null}$ 
\If{discard}
    \State $\Call{JobQueue.empty}{\null}$
\EndIf
\State $\Call{JobQueue.push}{jobid}$
\State extranonce2 $\gets$ $\Call{random}{extranonce2\_size}$
\State target $\gets \Call{getTarget}{nbits}$
\Repeat
    \State coinbase $\gets$ coin1 + extranonce1 + extranonce2 + coin2 
    \State merkle\_root $\gets \Call{doubleSHA2}{coinbase}$
    \ForEach {hash $\in$ merkle\_branch }
        \State merkle\_root $\gets \Call{doubleSHA2}{merkle\_root + hash}$
    \EndFor
    \State blockheader = ver + prevhash + $\Call{reverse}{merkle\_root}$ + ntime + nbits
    \State nonce $\gets 0$
    \For{nonce $\gets 0$ to $2^{32}-1$}
    \State blockhash $\gets \Call{doubleSHA2}{blockheader+nonce}$ 
    \If{blockhash $\leq$ target}
        \State $\Call{mining.submit}{username, jobid, extranonce2, ntime, nonce}$
        \State \Break
    \EndIf
    \EndFor
    \State extranonce2++
\Until{blockhash $\leq$ target}
\end{algorithmic}
\end{algorithm}

\section{Experimental Results}
We benchmarked the mining algorithm on four separate platforms with different conditions- 
\begin{enumerate}
    \item A modern 64-bit Processor
    \item A PSP Emulator
    \item A PSP-2000 system
    \item An ESP32 Embedded system
\end{enumerate}
For benchmarking purposes, we generated a fake blockheader with a target of 0, thus making the target impossible to reach, then we calculated the hash rate using the execution time for different number of iterations. The number of iterations were doubled after each run.

\subsection{64-bit Processor}
We ran the experiment on an Intel Core i5 8265U with 8 GB of RAM. 
The benchmarking program was compiled using gcc version 8.1.0 with no extra compiler flags. Both the naive and the optimized SHA-256 implementations were tested and compared. The CPU utilization was around 28-29\% for the duration of the test, while the RAM utilization was 0.4 MB throughout. 
The experiment was repeated 4 times. For some small number of iterations, sometimes the execution times was recorded as 0 due to the execution being too fast, in those cases the values were ignored for the average, if the value was 0 for all trials, then the data point was removed. 

The results of the test are tabulated in Table \ref{tab:64_results}.
\renewcommand\theadfont{}
\begin{table}[h]
    \centering
        \begin{tabular}{rll}
        \toprule
        \addlinespace
             & \multicolumn{2}{c}{\makecell{Average Hashrate (H/s) }} \\
     \cmidrule(lr){2-3}\thead{Number of \\iterations}&\thead{Naive}&\thead{Optimized}
    \tabularnewline 
        \midrule
                           16 &                        - &                     15000.00 \\
                           32 &                 31000.00 &                     31000.00 \\
                           64 &                 63000.00 &                     63000.00 \\
                          128 &                105833.33 &                    127000.00 \\
                          256 &                 77916.67 &                    255000.00 \\
                          512 &                115177.78 &                    383250.00 \\
                         1024 &                120720.09 &                    253618.75 \\
                         2048 &                251567.33 &                    827329.16 \\
                         4096 &                188053.53 &                    374599.43 \\
                         8192 &                214270.29 &                    441157.30 \\
                        16384 &                210652.88 &                    383179.96 \\
                        32768 &                194289.33 &                    338385.94 \\
                        65536 &                201467.11 &                    369246.27 \\
                       131072 &                202249.55 &                    395596.56 \\
                       262144 &                193006.37 &                    369746.15 \\
                       524288 &                195606.63 &                    381831.79 \\
                      1048576 &                161461.30 &                    377883.33 \\
                      2097152 &                166913.84 &                    470727.05 \\
                      4194304 &                173939.69 &                    468077.87 \\
                      8388608 &                190644.65 &                    478027.26 \\
                     16777216 &                186388.04 &                    470148.65 \\
        \bottomrule
        \end{tabular}
    \caption{Experimental Results on x64 processor}
    \label{tab:64_results}
\end{table}

\begin{figure}[h]
\centering
\includegraphics[width=0.45\textwidth]{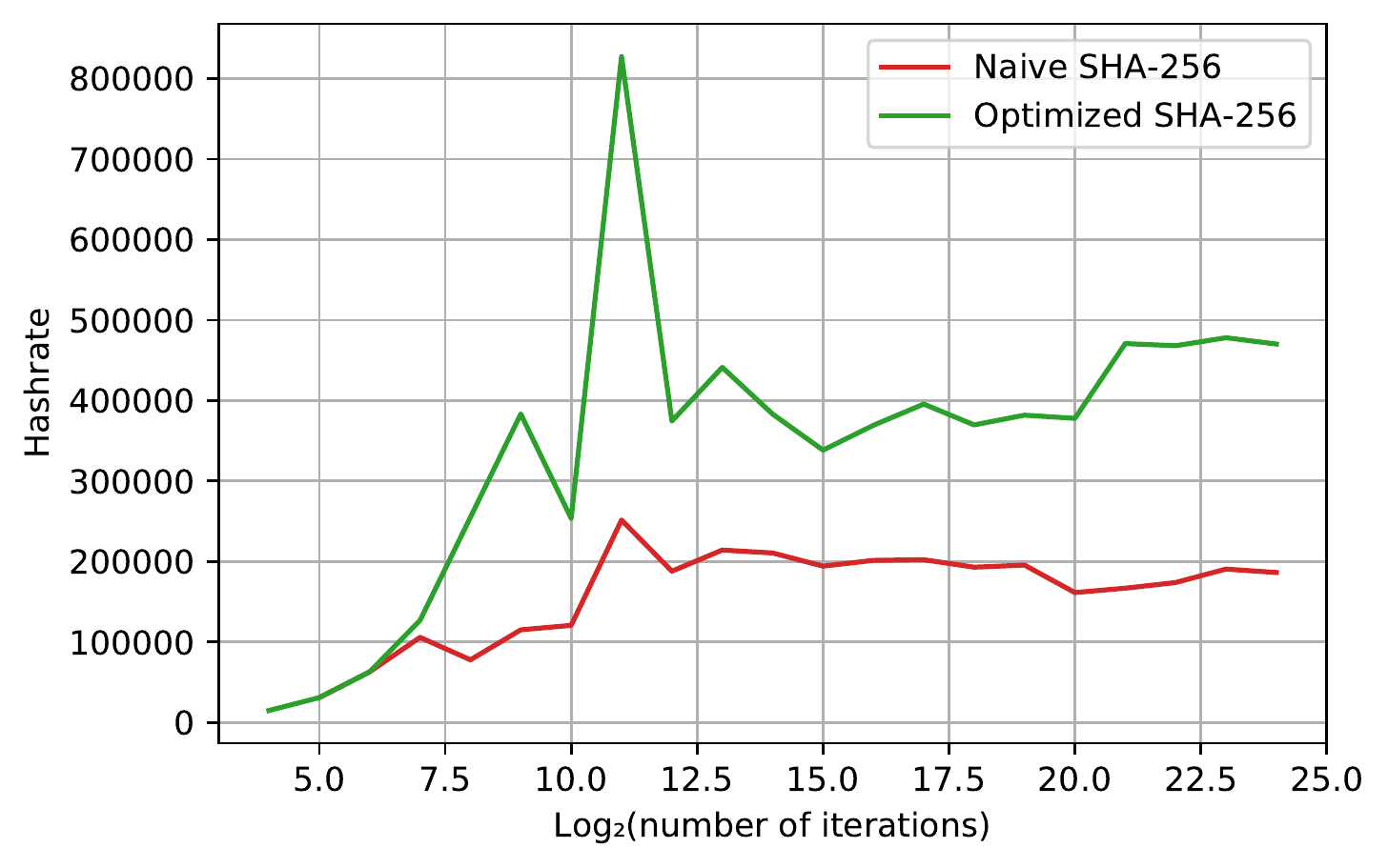}
\caption{Results of Benchmark on PC}
\label{fig:h_vs_i}
\end{figure}

We plotted the the hash rate with respect to the the Log$_2$ of the number of iterations (Figure \ref{fig:h_vs_i}), since the number of iterations were doubled after each run.

The average hashrate using naive algorithm was 162207.92 H/s, the maximum hashrate was 251567.33 H/s and the minimum was 31000 H/s. For the optimized algorithm, the average hashrate was 346371.69 H/s, the maximum hashrate was 827329.16 H/s and the minimum was 15000 H/s. On average the optimized algorithm was 2.17 times faster than the naive approach. In the best case scenario the optimized algorithm ran 3.27 times faster. In the worst case scenario the performance was equal.

\subsection{Emulator}
We ran the experiment on the PPSSPP Emulator \cite{ppsspp}.

The results of the test are tabulated in Table \ref{tab:emu_results}.

\renewcommand\theadfont{}
\begin{table}[h]
    \centering
        \begin{tabular}{rll}
        \toprule
        \addlinespace
             & \multicolumn{2}{c}{\makecell{Average Hashrate (H/s) }} \\
     \cmidrule(lr){2-3}\thead{Number of \\iterations}&\thead{Naive}&\thead{Optimized}
    \tabularnewline 
        \midrule
               2 &   3557.8614 &   3676.4706 \\
               4 &   1834.8624 &   5545.2865 \\
               8 &   2754.8209 &   6465.4857 \\
              16 &   3216.9118 &   6944.4444 \\
              32 &   3448.2759 &   7179.2497 \\
              64 &   3285.2904 &   6722.8684 \\
             128 &   3585.8615 &   7346.9860 \\
             256 &   3556.1280 &   7223.7960 \\
             512 &   3568.4299 &   7234.3739 \\
            1024 &   3553.5466 &   8196.6233 \\
            2048 &   3563.7149 &   7217.7598 \\
            4096 &   3542.0495 &   7178.9208 \\
            8192 &   3559.8976 &   7154.5116 \\
           16384 &   3557.0436 &   7177.0398 \\
           32768 &   3555.6086 &   7173.5082 \\
           65536 &   3557.6478 &   7167.1899 \\
          131072 &   3558.4293 &   7172.5261 \\
          262144 &   3559.6915 &   7173.3474 \\
          524288 &   3561.3188 &   7175.0318 \\
         1048576 &   3560.4792 &   7176.3186 \\
        \bottomrule
        \end{tabular}
    \caption{Experimental Results on the Emulator}
    \label{tab:emu_results}
\end{table}

\begin{figure}[h]
\centering
\includegraphics[width=0.45\textwidth]{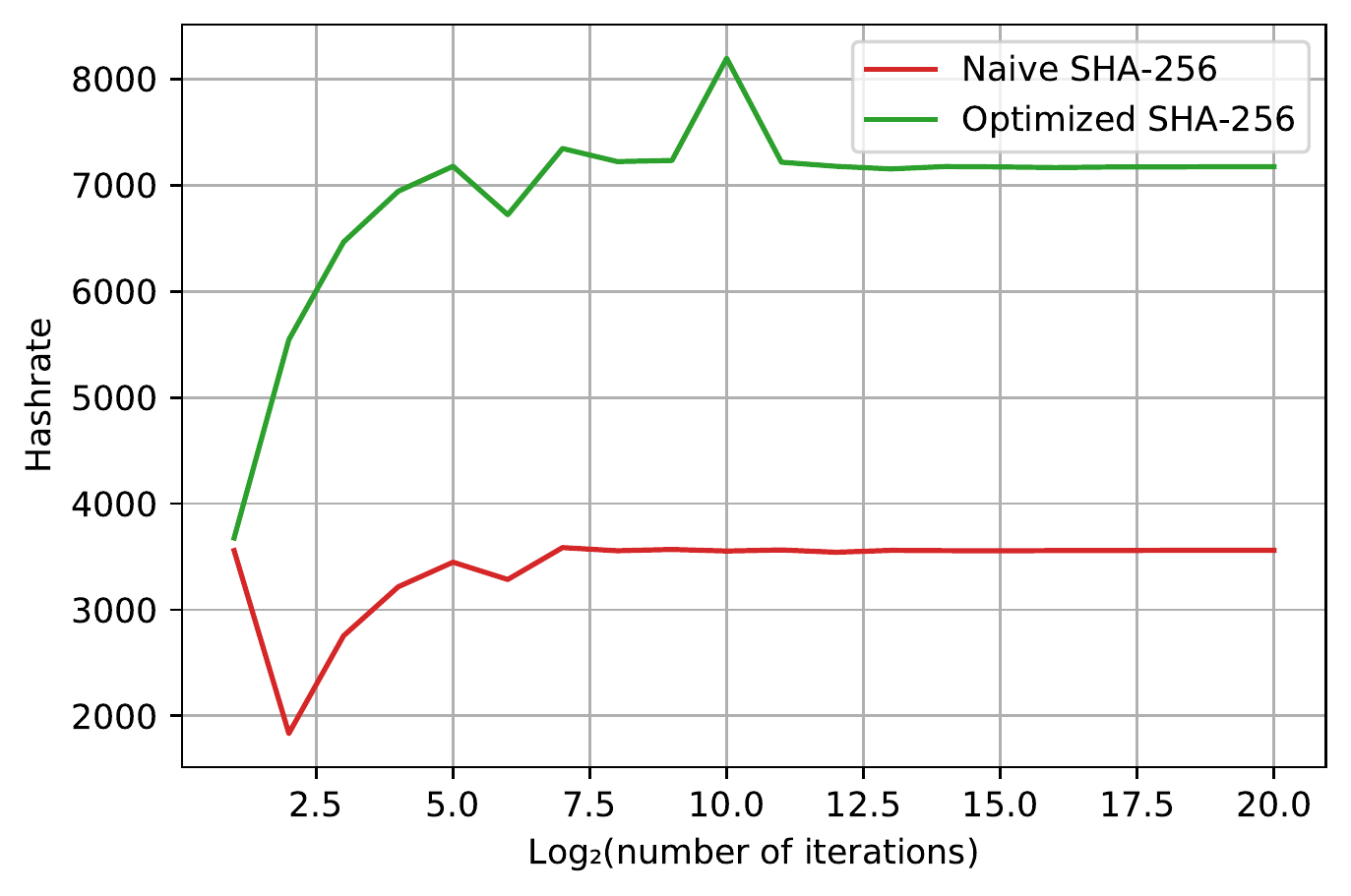}
\caption{Results of Benchmark on the Emulator}
\label{fig:emu_h_vs_i}
\end{figure}

The data follows the same general trends as the 64-bit processor.

The average hashrate using naive algorithm was 3396.89 H/s, the maximum hashrate was 3585.86 H/s and the minimum was 1834.86 H/s. For the optimized algorithm, the average hashrate was 6915.08 H/s, the maximum hashrate was 8196.62 H/s and the minimum was 3676.47 H/s. On average the optimized algorithm was 2.06 times faster than the naive approach. In the best case scenario the optimized algorithm ran 3.02 times faster. In the worst case scenario the performance was approximately equal.

The results were much more consistent, compared to the PC run.

\subsection{PSP-2000}

We ran the experiment on PSP-2000 with the processor's clockspeed set to 333 Mhz.

\renewcommand\theadfont{}
\begin{table}[h]
    \centering
        \begin{tabular}{rll}
        \toprule
        \addlinespace
             & \multicolumn{2}{c}{\makecell{Average Hashrate (H/s) }} \\
     \cmidrule(lr){2-3}\thead{Number of \\iterations}&\thead{Naive}&\thead{Optimized}
    \tabularnewline 
        \midrule
                  2 &   1615.5089 &   3003.0030 \\
                  4 &   2461.0336 &   4622.4961 \\
                  8 &   2887.7888 &   5464.4809 \\
                 16 &   3099.8140 &   5870.8415 \\
                 32 &   3174.2781 &   6079.6235 \\
                 64 &   3243.9112 &   6119.4755 \\
                128 &   3271.6781 &   6205.7271 \\
                256 &   3283.3323 &   6223.9145 \\
                512 &   3292.1864 &   6244.5009 \\
               1024 &   3294.9831 &   6255.1209 \\
               2048 &   3297.2148 &   6253.5132 \\
               4096 &   3297.9007 &   6259.0083 \\
               8192 &   3298.3937 &   6259.5621 \\
              16384 &   3298.7126 &   6260.1475 \\
              32768 &   3298.7734 &   6260.4605 \\
              65536 &   3298.8467 &   6260.6111 \\
             131072 &   3298.8771 &   6260.6526 \\
             262144 &   3298.8890 &   6260.6793 \\
             524288 &   3298.8936 &   6260.7004 \\
            1048576 &   3298.8996 &   6260.7034 \\
        \bottomrule
        \end{tabular}
    \caption{Experimental Results on PSP-2000}
    \label{tab:psp_results}
\end{table}

\begin{figure}[h]
\centering
\includegraphics[width=0.45\textwidth]{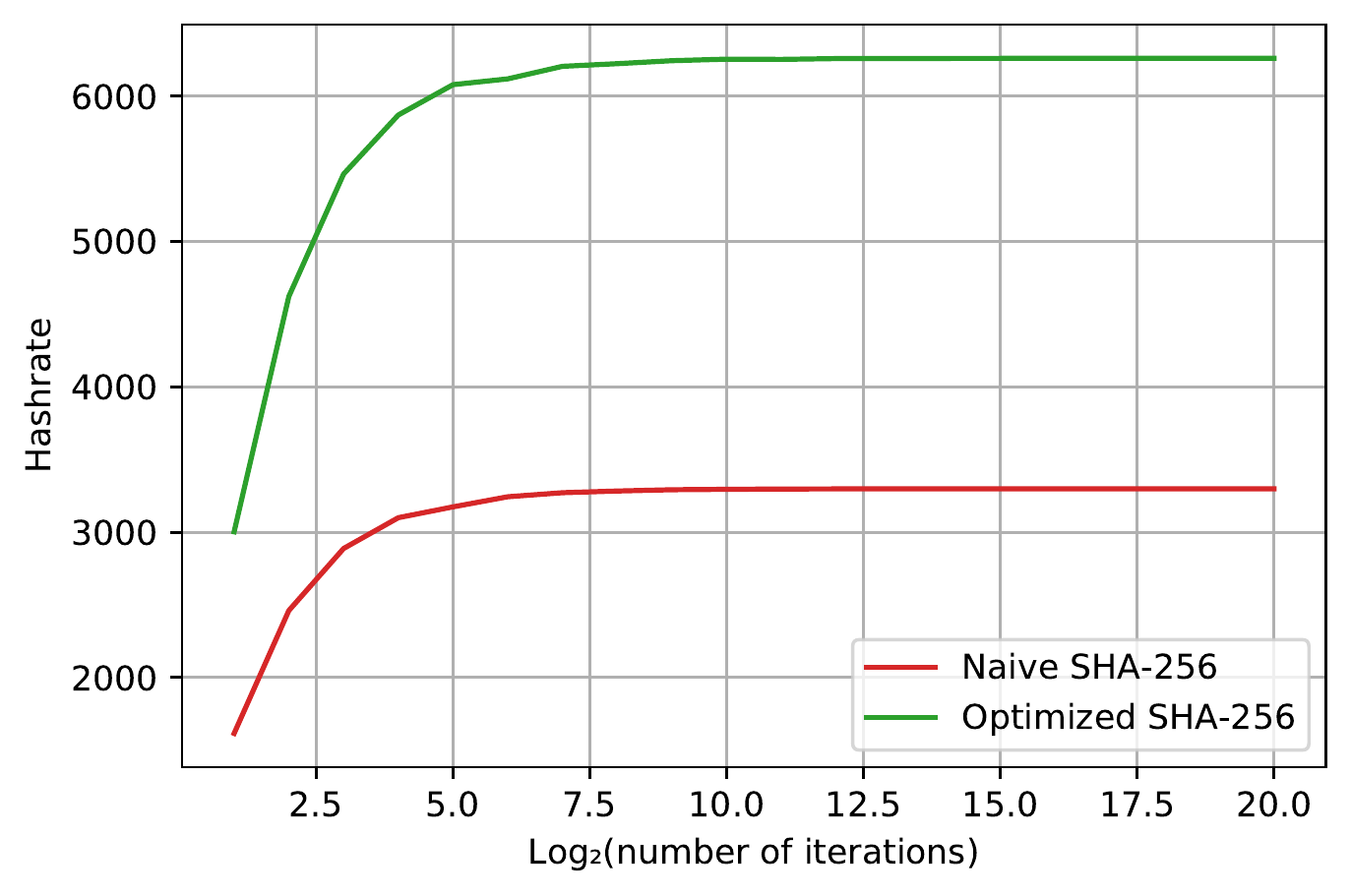}
\caption{Results of Benchmark on PSP-2000}
\label{fig:psp_h_vs_i}
\end{figure}

The results of the test are tabulated in Table \ref{tab:psp_results}.

Once again data follows the same general trends as the 64-bit processorand the emulator.

The average hashrate using naive algorithm was 3396.89 H/s, the maximum hashrate was 3585.86 H/s and the minimum was 1834.86 H/s. For the optimized algorithm, the average hashrate was 6915.08 H/s, the maximum hashrate was 8196.62 H/s and the minimum was 3676.47 H/s. On average the optimized algorithm was 2.06 times faster than the naive approach. In the best case scenario the optimized algorithm ran 3.02 times faster. In the worst case scenario the performance was approximately equal.

The results were even more consistent than the emulator. The average hashrate was around 10\% lower than the emulator and around 50 times lower than the PC which is to be expected since there is a huge power difference, not to mention that the the microprocessor is almost 30 years old. However, for a processor that came out 20 years before Bitcoin was even developed, the hashrate is higher than expected and might be useful in certain scenarios.

\subsection{ESP32}
\renewcommand\theadfont{}
\begin{table}[h]
    \centering
        \begin{tabular}{rll}
        \toprule
        \addlinespace
             & \multicolumn{2}{c}{\makecell{Average Hashrate (H/s) }} \\
     \cmidrule(lr){2-3}\thead{Number of \\iterations}&\thead{Naive}&\thead{Optimized}
    \tabularnewline 
        \midrule
              2 &   7299.2701 &   8928.5714 \\
              4 &   9677.4194 &  13157.8947 \\
              8 &  11513.1579 &  13565.8915 \\
             16 &  11450.3817 &  13513.5135 \\
             32 &  11443.3370 &  13531.2091 \\
             64 &  11446.2209 &  13539.6518 \\
            128 &  11456.9238 &  13533.6743 \\
            256 &  11455.0110 &  13530.0048 \\
            512 &  11457.1422 &  13533.5558 \\
           1024 &  11458.3333 &  13532.2830 \\
           2048 &  11458.4791 &  13531.9160 \\
           4096 &  11458.1351 &  13531.7774 \\
           8192 &  11458.3158 &  13531.9093 \\
          16384 &  11458.3100 &  13531.9417 \\
          32768 &  11458.3271 &  13531.9132 \\
          65536 &  11458.3237 &  13531.8738 \\
         131072 &  11458.3330 &  13531.9002 \\
         262144 &  11458.3316 &  13531.8868 \\
         524288 &  11458.3334 &  13531.8983 \\
        1048576 &  11458.3332 &  13531.8967 \\
        2097152 &  11458.3334 &  13531.8944 \\
        4194304 &  11458.3333 &  13531.8934 \\
        8388608 &  11458.3333 &  13531.8922 \\
        \bottomrule
        \end{tabular}
    \caption{Experimental Results on ESP32}
    \label{tab:esp32_results}
\end{table}

\begin{figure}[h]
\centering
\includegraphics[width=0.45\textwidth]{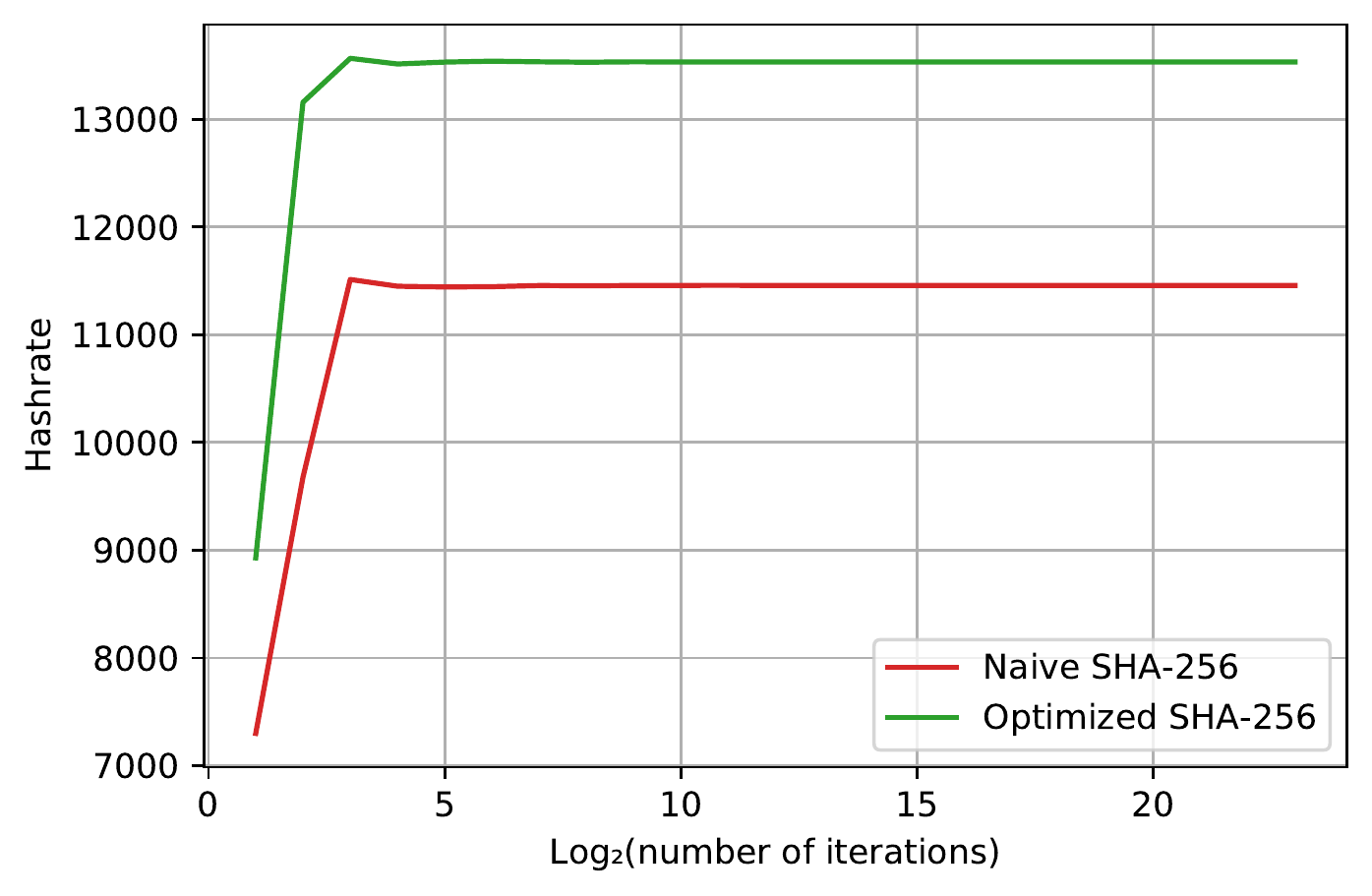}
\caption{Results of Benchmark on ESP32}
\label{fig:esp32_h_vs_i}
\end{figure}

The results of the test are tabulated in Table \ref{tab:esp32_results}.

The average hashrate using naive algorithm was 11200.67 H/s, the maximum hashrate was 11513.16
 H/s and the minimum was 7299.27 H/s. For the optimized algorithm, the average hashrate was 13316.56 H/s, the maximum hashrate was 13565.89 H/s and the minimum was 8928.57 H/s. On average the optimized algorithm was 1.19 times faster than the naive approach. In the best case scenario the optimized algorithm ran  1.36 times faster. In the worst case scenario the optimized algorithm was 1.18 faster.

Surprisingly the difference between the naive and the optimized was very small compared  to all other platforms, the results were consistent after the small disturbance in the start. Even though the hardware is much newer it's only about twice as fast as the PSP-2000. 

\begin{figure}[h]
\centering
\includegraphics[width=0.45\textwidth]{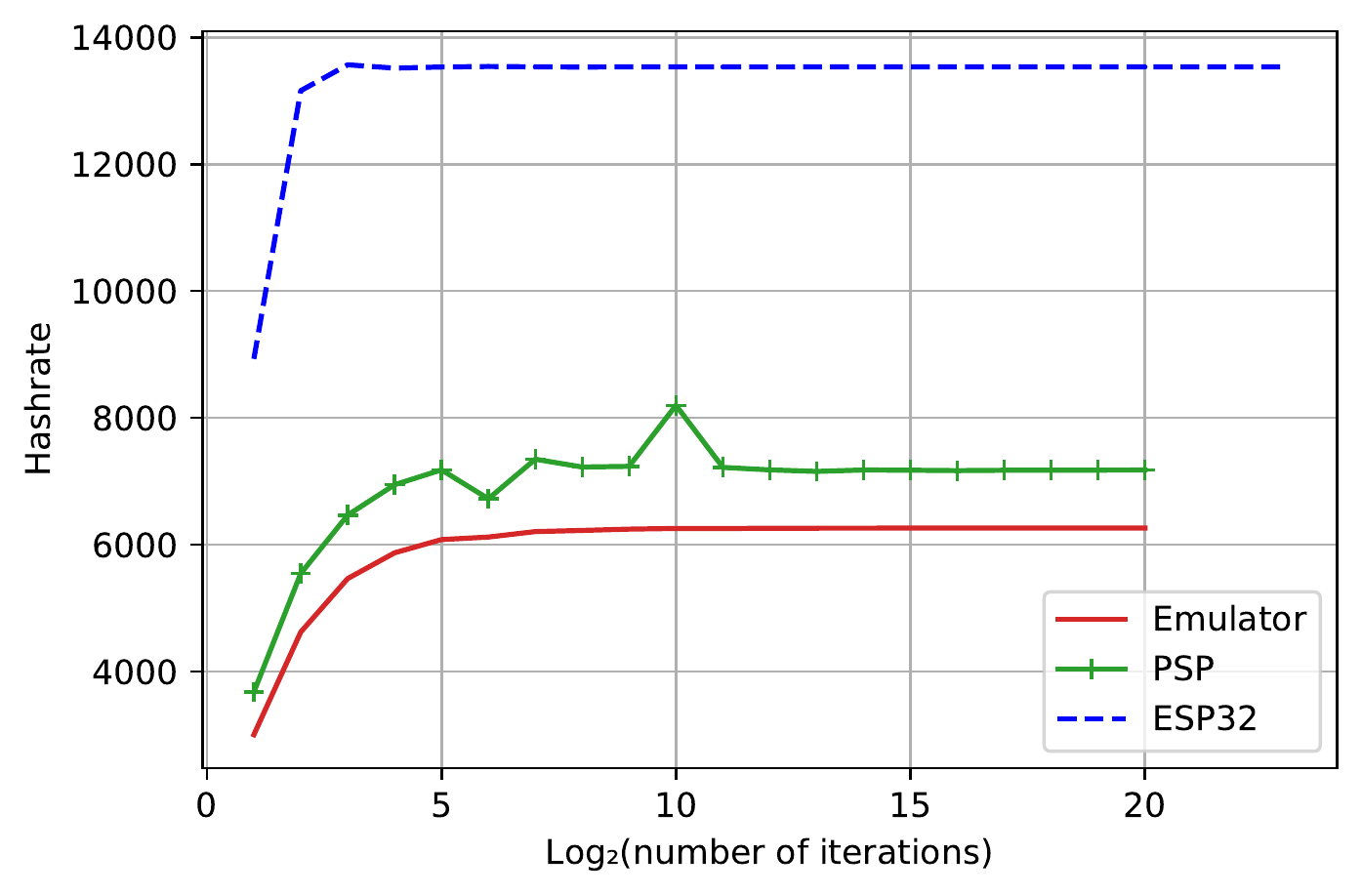}
\caption{Comparison of all the implemented platforms}
\label{fig:all_h_vs_n}
\end{figure}

\section{Conclusion}
This research represents the first published work in literature to mine traditional cryptocurrencies like Bitcoin on low-power IoT devices without the need of connecting to an external device. Our algorithm is platform independent and can be easily ported to most IoT systems, it makes no assumptions about the hardware, including the support for the C standard library. To illustrate this we implemented the algorithms on both an old unsupported hardware and a modern low-cost IoT device. In both cases we were successfuly able to mine Bitcoins as well as connect to the Bitcoin blockchain. To make it as efficient as possible, we optimized the SHA-256 algorithm reaching almost twice the hashrate of the naive algorithm in most cases. Results were very consistent, the PSP was able to reach a maximum hashrate of 8196.62 H/s while the ESP32 was able to reach 13565.89 H/s. The source code is made publicly available with detailed instructions so that it can be ported to any platform. 
\clearpage

\bibliographystyle{elsarticle-num} 
\bibliography{cas-refs}
 
\end{document}